\documentclass[manuscript]{aastex6}

\usepackage{epsfig}
\usepackage{graphicx}
\usepackage{amssymb}
\usepackage{amsmath}
\usepackage{times}
\usepackage{hyperref}
\usepackage{natbib}
\usepackage{blindtext}
\usepackage{roussev}
\usepackage{color}
\usepackage{mathrsfs}

\newcount\doepsf
\newcommand{\alf}{Alfv\'en }

\newcommand{\be}{\begin{equation}}
\newcommand{\ee}{\end{equation}}
\newcommand{\bea}{\begin{eqnarray}}
\newcommand{\eea}{\end{eqnarray}}
\newcommand{\bean}{\begin{eqnarray*}}
\newcommand{\eean}{\end{eqnarray*}}

\begin{document}
\title{Toward  Quantitative Model for Simulation and Forecast of Solar Energetic Particles Production during Gradual Events - II: Kinetic Description of SEP}

\author{
D. Borovikov,     \altaffilmark{1,2}
I.~V. Sokolov,    \altaffilmark{1}
Z. Huang,
\altaffilmark{1}
I.~I. Roussev,    \altaffilmark{3}
and T.~I. Gombosi,\altaffilmark{1}
}

\altaffiltext{1}{Center for Space Environment Modeling, University of Michigan,
2455 Hayward St, Ann Arbor, MI 48109; \\dborovik@umich.edu, igorsok@umich.edu,
tamas@umich.edu.}
\altaffiltext{2}{Space Science Center, University of New Hampshire, 8 College Road Durham, NH 03824}
\altaffiltext{3}{National Science Foundation; iroussev@nsf.gov}
\begin{abstract}
Solar Energetic Particles (SEPs) possess a high destructive potential as 
they pose multiple radiation hazards on Earth and onboard spacecrafts.
The present work continues a series started with the paper by \citet{boro18} describing a computational tool to simulate and, potentially, predict the SEP threat based on the observations of the Sun.
Here we present the kinetic model coupled with the global MHD model for the Solar Corona (SC) and Inner Heliosphere (IH),
which was described in the first paper in the series. 
At the heart of the coupled model is a self-consistent treatment of the \alf wave turbulence. 
The turbulence not only heats corona, powers and accelerates the solar wind, but also serves as the main agent to scatter the SEPs and thus controls their acceleration and transport. 
The universal character of the turbulence in the coupled model provides a realistic description of the SEP transport by using the level of turbulence as validated with the solar wind and coronal plasma observations. At the same time, the SEP observations at 1 AU can be used to validate the model for turbulence in the IH, since the observed SEPs have witnessed this turbulence on their way through the IH. 
\end{abstract}


\numberwithin{equation}{section}


\section{Introduction}
\label{sec:kinetic:intro}
The kinetic transport of energetic particle population 
through the inter-planetary space is an important problem in space science.
It was studied since the discovery of Galactic Cosmic Rays (GCR),
energetic particles originating
from beyond the Solar system.
A comprehensive summary of the problem can be found in the review by
\citet{Parker1965}.
Although results in the said review are 
obtained in a different context,
some can be readily applied for SEP transport.

In the present paper we discuss the numerical methods and tools to solve the realistic kinetic equations in application to the solar energetic particle acceleration and transport.

The present paper is structured as follows.
In Section~\ref{sec:basic} we state the basic concepts that 
both theoretical and numerical aspects of 
the kinetic description of SEP rely upon.

\section{Basic concepts}
\label{sec:basic}

\subsection{SEP distribution function}
\label{sec:basic:distr}
As SEP population forms a suprathermal tail
of particle distribution in the solar wind,
we need to start developing the kinetic treatment
from a distribution function of a  general form. 
We characterize SEPs by a (canonical) distribution function 
$F({\bf R},{\bf p},t)$ of coordinates, ${\bf R}$, and momentum, ${\bf p}$, 
as well as time, $t$, such that the number of particles, $dN$, 
within the elementary volume,  $d^3{\bf R}$, 
is given by the following normalization integral: 
$dN=d^3{\bf R}\int{d^3{\bf p}\,F({\bf R},{\bf p},t)}$. 
In a magnetized moving plasma, 
it is convenient to consider the distribution function at any given point, 
${\bf R}$, in the co-moving frame of reference, 
which moves with the local plasma velocity, 
${\bf u}({\bf R},t)$.
Also, we introduce spherical coordinates, 
$(p=|{\bf p}|,\,\mu={\bf b}\cdot{\bf p}/p,\,\varphi)$,
in the momentum space with its polar axis aligned with the direction, ${\bf b}=\mathbf{B}/B$, 
of the magnetic field, $\mathbf{B}({\bf R},t)$.
Herewith, $\mu$ is the cosine of pitch-angle.
The normalization integral in these new variables becomes: 
\mbox{$dN=d^3{\bf R}\int_0^\infty{p^2dp\int_{-1}^1{d\mu\int_0^{2\pi}{d\varphi\,F({\bf R}, p,\mu,\varphi,t)}}}$}. 

Using the canonical distribution function, 
one can also define a gyrotropic distribution function, 
$f({\bf R}, p,\mu,t)=\frac1{2\pi}\int_0^{2\pi}{d\varphi \,F({\bf R}, p,\mu,\varphi,t)}$. 
This function is designed to describe the particle motion averaged over 
the phase of  gyration about the magnetic field.  
The isotropic (omnidirectional) distribution function, 
$f_0({\bf R}, p,t)=\frac1{2}\int_{-1}^1{d\mu\, f({\bf R}, p,\mu,t)}$ 
is additionally averaged over pitch angle. 
The normalization integrals are: 
$dN=2\pi d^3{\bf R}\int_0^\infty{p^2dp\int_{-1}^1{d\mu \,f({\bf R}, p,t)}}=4\pi d^3{\bf R}\int_0^\infty{p^2dp \,f_0({\bf R}, p,t)}$
The kinetic equation for the isotropic part of the distribution function, $f_0\left(\mathbf{R},p,t\right)$, was introduced in \citet{Parker1965}:
\begin{equation}
\label{eq:parker}
\frac{\partial f_0}{\partial t} + 
\left(\mathbf{u}\cdot\nabla\right)f_0 -
\frac{1}{3}\left(\nabla\cdot\mathbf{u}\right)
\frac{\partial f_0}{\partial\ln p} = 
\nabla\cdot\left(\varkappa\cdot\nabla f_0\right) + S,
\end{equation}
where $\varkappa=D_{xx}\mathbf{b}\mathbf{b}$ 
is the tensor of parallel (spatial) diffusion along the magnetic field, 
$S$ is the source term.
In this approximation,  the cross-field diffusion of particles is neglected.
The Parker Eq.~\ref{eq:parker} captures effects that Interplanetary Magnetic Field (IMF) and other background parameters of the solar wind 
on the SEP transport and acceleration.
The term proportional to the divergence of $\mathbf{u}$ is the 
adiabatic cooling, for $\left(\nabla\cdot\mathbf{u}\right)>0$,  
or (the first order Fermi) acceleration in compression or shock waves.
In the companion paper \cite{boro18} we provided preliminary results for the SEP acceleration and transport obtained by solving Eq.~\ref{eq:parker} numerically.

\subsection{Flux/Lagrangian Coordinates }
\label{sec:basic:Lagr}
Our model of SEP transport and acceleration is
based on the assumption that particles don't decouple
from their field lines. 
In other words, we assume that 
particle motion in physical space consists of: 
(a) displacement of particle's guiding center along some IMF line; and 
(b) joint advection of both the guiding center and the IMF line 
together with plasma
into which the field is frozen. 
Mathematically, the method employs Lagrangian coordinates, ${\bf R}_L$, 
which stay with advecting fluid elements rather 
than with fixed positions in space. 
As each fluid element moves, its Lagrangian coordinates, ${\bf R}_L$, 
remain unchanged,
while its spatial location, ${\bf R}\left({\bf R}_L,t\right)$, 
changes in time in accordance with the local velocity of plasma,~$\mathbf{u}(\mathbf{R},t)$:
\begin{equation}
\label{eq:DxDt}
\frac{D{\bf R}({\bf R}_L,t)}{Dt}={\bf u}({\bf R},t)
\end{equation}
Herewith, the partial time derivative at constant Lagrangian coordinates (also referred to as substantial derivative), 
${\bf R}_L$, is denoted as $\frac{D}{Dt}$, while the notation 
$\frac\partial{\partial t}$ is used to denote the partial time derivative at constant Eulerian coordinates, ${\bf R}$. 
The two are related as 
$\frac{D}{Dt}=\frac{\partial}{\partial t}+\mathbf{u}\cdot\nabla$.

Certain terms in Parker Eq.~\ref{eq:parker} as well
as other equations considered in this paper may be
expressed in term of the Lagrangian derivatives and spatial derivative along lines ($\partial / \partial s = \textbf{b} \cdot \nabla$) using equations of the plasma motion. 
Particularly, the continuity equation for the plasma density, $\rho(\mathbf{R},t)$, can be represented as follows:
\begin{equation}\label{eq:Lagr:continuity}
\nabla\cdot\mathbf{u}=-\frac{D\ln\rho}{Dt}
\end{equation}
Expressing the induction equation using the substantial derivative, 
$\frac{D}{Dt}$, we obtain
\be
\label{eq:Lagr:induction}
\left(\mathbb{I}-\mathbf{bb}\right):\nabla\mathbf{u} = - \frac{D \ln{B}}{D t}
\ee
where $\rho$ is the plasma density,
$\mathbb{I}$ is the identity matrix.
The time-dependent
changes in the distance between two neighboring Lagrangian meshes, $\delta s$,
is described by the following evolutionary equation \cite[e.g.][]{Landau1959}:
\bea
\label{eq:Lagr:distance}
\frac{D \ln{\delta s}}{D t} =\mathbf{bb}:\nabla\mathbf{u}
\eea

Using Eqs.~\ref{eq:Lagr:continuity},~\ref{eq:Lagr:induction} the latter can be also written as:
\begin{equation}
\label{eq:Lagr:distance:Brho}
\frac{D \ln{\delta s}}{D t} =\frac{D\ln(B/\rho)}{Dt}
\end{equation}
Eq.~\ref{eq:Lagr:distance:Brho} may be applied 
to derive relation between Lagrangian and Eulerian distances, 
$s_L$ and $s$. 
With the initial condition $\frac{\partial s}{\partial s_L}=1$ at $t=0$ we have:
\begin{equation}
\frac{\partial s(s_L,t)}{\partial s_L}=\frac{B(s_L,t)\rho(s_L,0)}{B(s_L,0)\rho(s_L,t)}       
\end{equation}

From the solenoidal constraint, $\nabla \cdot \textbf{B} = 0$, one can also find that: 
\be
\label{eq:divb}
\nabla \cdot \mathbf{b} = -
\frac{\ln{B}}{\partial s}
\ee

We apply the formalism presented above whenever possible.
For example, Parker equation may be rewritten as follows:
\begin{equation}
    \label{eq:parker:Lagr}
    \frac{Df_0}{Dt}+\frac13\frac{D\ln\rho}{Dt}\frac{\partial f_0}{\partial\ln p}=
    \nabla\cdot\left(\varkappa\cdot\nabla f_0\right)+S
\end{equation}
Such formulation of mathematical problem is
particularly convenient for translating it into
a numerical model and has been used in development of M-FLAMPA (see Section~\ref{sec:mflampa}). 

\section{Focused Transport Equation and Diffusive Limit}
\label{sec:kinetic:FTE}

As mentioned in Section~\ref{sec:basic:distr} the Parker equation
captures major effects that the background has on SEP population.
The Parker equation was used to develop a Diffuse Shock Acceleration (DSA) theory
\citep{axford77,krymsky77,Bell1978a,Bell1978b,blandford78,axford81}, which predicts the power-law spectrum of galactic cosmic rays, close to the observed one.
However, due to pitch-angle dependant part of 
the distribution function not being featured in the Parker 
equation explicitly, its importance to formulation of DSA 
may be lost in the context.
In this section we consider a distribution
function $f\left(\mathbf{R}, p,\mu,t\right)$
and demonstrate how diffusive behavior derives
from its pitch-angle-dependant part.

When pitch angles of particles are taken into account,
one needs to consider the appropriate scattering in the 
momentum space.
The equation 
for a non-relativistic gyrotropic distribution function
$f\left(\mathbf{R},p,\mu,t\right)$
can be found in, for example,  \citet{Skilling1971}:
\bea
\label{eq:skilling}
\frac{\partial f}{\partial t} &+& 
\left(\mathbf{u}+\mu v\mathbf{b}\right)\cdot
\nabla f + \nonumber\\
&+& \left[
\frac{1-3\mu^2}{2}\left(\mathbf{bb}:\nabla\mathbf{u}\right)-
\frac{1-\mu^2}{2}\left(\nabla\cdot\mathbf{u}\right) - 
\frac{\mu}{v}\left(\mathbf{b}\cdot\frac{D\mathbf{u}}{Dt}\right)
\right]\frac{\partial f}{\partial\ln p} +
\\
&+& \frac{1-\mu^2}{2}\left[
v\left(\nabla\cdot\mathbf{b}\right)
-3\mu\left(\mathbf{bb}:\nabla\mathbf{u}\right)
+ \mu\left(\nabla\cdot\mathbf{u}\right)-
\frac{2}{v}\left(\mathbf{b}\cdot\frac{D\mathbf{u}}{Dt}\right)
\right]
\frac{\partial f}{\partial\mu}=
\frac{\delta f}{\delta t} + S,\nonumber
\eea

The particle scattering rate, $\frac{\delta f}{\delta t}$, 
in this model is due to the particle interaction with the \alf wave turbulence. An important physical effect related to particles' pitch angle distribution
is the focusing effect \citep[][and references therein]{earl76},
also referred to as focused transport.
The effect takes places under conditions that constrain pitch-angle
scattering across $\mu{=}0$.
In the extreme case, when particles can't change the direction of their
propagation along their field lines, the whole population is effectively
split into two independent hemispheric subpopulations,
one of particles propagating inward, the other of particles
propagating outward.
The implications of such splitting have been explored in \citet{Isenberg1997}.
Effects of interaction of particles with solar wind plasma and IMF
such as adiabatic cooling/heating on the focused transport have been studied
in, for example, \citet{Ruffolo1995}.
A detailed view on different aspects of evolution of distribution
of particles propagating along magnetic-field lines, 
i.e. convection, cooling/heating, and focusing,
can be found in \citet{Kota:1997}.


Relations in Eqs.~\ref{eq:Lagr:continuity},~\ref{eq:Lagr:induction}~and~\ref{eq:divb} allow a new treatment of Eq.~\ref{eq:skilling}
for acceleration and field-aligned transport of SEPs 
\citep{Kota:2004,Kota2005}:
\bea
\label{eq:kota}
\frac{Df}{Dt} &+& 
v\mu\frac{\partial f}{\partial s} +\left[
\frac13\frac{D\ln\rho}{Dt}+
\frac{1-3\mu^2}{6}\frac{D\ln \left(B^3/\rho^2\right)}{Dt}
-\frac{\mu}{v}\mathbf{b}\cdot\frac{D\mathbf{u}}{Dt}
\right]p\frac{\partial f}{\partial p}+\nonumber\\
&+& \frac{1-\mu^2}{2}\left[
-v\frac{\partial\ln B}{\partial s} +
\mu\frac{D\ln\left(\rho^2/B^3\right)}{Dt}- 
\frac{2}{v}\mathbf{b}\cdot\frac{D\mathbf{u}}{Dt}
\right]\frac{\partial f}{\partial\mu}=
\left(\frac{\delta f}{\delta t}\right)_{\rm scat}
\eea
Or, in the conservative form, which is convenient both for solving the equation numerically using the conservative scheme and for analytical derivations:
\bea
\label{eq:kotacons}
\frac{Df}{Dt}&+& vB\frac{\partial}{\partial s}\left[\frac{\left(1-\mu^2\right)}{2B}\frac{\partial f}{\partial\mu}\right]+\frac1{p^2}\frac\partial{\partial p}\left\{p^3\left[
\frac{1-3\mu^2}{6}\frac{D\ln \left(B^3/\rho^2\right)}{Dt}
-\frac{\mu}{v}\mathbf{b}\cdot\frac{D\mathbf{u}}{Dt}
\right]f\right\}+ \\
&+& \frac{\partial }{\partial\mu}\left\{\frac{1-\mu^2}2\left[\left(
\mu\frac{D\ln\left(\rho^2/B^3\right)}{Dt}- 
\frac{2}{v}\mathbf{b}\cdot\frac{D\mathbf{u}}{Dt}\right)f-v\frac{\partial f}{\partial s}
\right]\right\}+\frac13\frac{D\ln \rho}{Dt}p\frac{\partial f}{\partial p} =
\left(\frac{\delta f}{\delta t}\right)_{\rm scat}\nonumber
\eea
The diffusive limit of Eq.~\ref{eq:kotacons} is less accurate but widely used. In general case, scattering integral may be applied \citep[see, e.~g.][]{sokolov06} in the Fokker-Planck form (note that $D_{\mu p}=D_{p\mu}$):
\bea
\label{eq_Fokker}
\left(\frac{\delta f}{\delta t}\right)_{\rm scat}=\frac1{p^2}\frac\partial{\partial p}\left[p^2\left(D_{pp}\frac{\partial f}{\partial p}+D_{p\mu}\frac{\partial f}{\partial \mu}\right)\right]+\frac\partial{\partial \mu}\left[D_{\mu p}\frac{\partial f}{\partial p}+D_{\mu\mu}\frac{\partial f}{\partial \mu}\right]
\eea
One can assume the particle speed to be large compared to the \alf speed, as well as the plasma speed $u \ll v$ , 
and suppose that $D_{\mu \mu}^{-1}$ is small compared
to any hydrodynamic time. 
Under these assumption one can 
treat the pitch-angle dependant part of 
the distribution function
as a small correction, $f_1$, 
to its isotropic part, $f_0$.
In other words, 
$f(\mathbf{R}, p, \mu, t) = f_0(\mathbf{R}, p, t) + f_1(\mathbf{R},p,\mu,t)$, 
where $f_1 \ll f_0$.
To obtain the evolutionary equation for $f_0$ 
(i.e. the Parker equation, Eq.~\ref{eq:parker}) let us average
Eq.~\ref{eq:kotacons} with respect to the particle pitch angle:
\bea
\label{eq_4_DSA}
\frac{D f_0}{D t} &+& B\frac{\partial}{\partial s}\left[\frac{v}{B}\left\langle\frac{\left(1-\mu^2\right)}{2}\frac{\partial f_1}{\partial\mu}\right\rangle_\mu\right] + \frac13 \frac{D \ln{\rho}}{D t} p \frac{\partial f_0}{ \partial p}=\nonumber\\&=& \frac1{p^2}\frac\partial{\partial p}\left[p^2\left(\left\langle D_{pp}\right\rangle_\mu\frac{\partial f_0}{\partial p}+\left\langle D_{p\mu}\frac{\partial f}{\partial \mu}\right\rangle_\mu\right)\right],
\eea
where $\langle ...\rangle_\mu =
\frac12 \int d \mu(...)$. The perturbation of the distribution function, $\partial f_1 / \partial \mu$, may be found by claiming that the flux along $\mu$ coordinate in Eq.~\ref{eq:kotacons} vanishes. Keeping in the expression for this flux only large terms, proportional to the scattering frequency or particle speed, we find:
\bea
\label{eq_3_DSA}
\frac{\partial f_1}{ \partial \mu} = - v \frac{1 - \mu^2}{
2 D_{\mu \mu}} \frac{\partial f_0} {\partial s}-\frac{D_{\mu p}}{D_{\mu\mu}}\frac{\partial f_0}{\partial p}.
\eea
The particle flux, $J$, may be found by averaging the parallel velocity, $$J=\left\langle \mu v f_1\right\rangle_\mu=v\left\langle\frac{\left(1-\mu^2\right)}{2}\frac{\partial f_1}{\partial\mu}\right\rangle_\mu=-D_{zz}\frac{\partial f_0}{\partial s}-\frac13\tilde{V}p\frac{\partial f_0}{\partial p},$$
with the following expressions for a spatial diffusion coefficient, $D_{zz}$, and average ion speed, $\tilde{V}$:
\be
\label{eq_6_DSA}
D_{zz}= v^2 \left\langle\frac{\left( 1 - \mu^2
\right)^2}{4 D_{\mu \mu}}\right\rangle_\mu, \qquad \tilde{V}=\frac{3v}{p}\left\langle\frac{\left(1-\mu^2\right)D_{\mu p}}{2D_{\mu\mu}}\right\rangle_\mu.
\ee
From here, we obtain the equation of the diffuse approximation, 
\bea
\label{eq_7_DSA}
\frac{D f_0}{D t} &+& B\frac{\partial}{\partial s}\left(\frac{J}{B}\right) + \frac{p}3 \frac{D \ln{\rho}}{D t} \frac{\partial f_0}{ \partial p}=
\frac1{p^2}\frac\partial{\partial p}\left[p^2\left(D_{F2}\frac{\partial f_0}{\partial p}-\frac{p}3\tilde{V}\frac{\partial f}{\partial s}\right)\right],\eea
where the second order Fermi acceleration coefficient is defined as $D_{F2}=\left\langle D_{pp}-\frac{D_{\mu p}^2}{D_{\mu\mu}}\right\rangle_\mu$. Eq.~\ref{eq_7_DSA} reduces to the Parker equation  \ref{eq:parker} in Lagrangian coordinates, if $D_{\mu p}=D_{pp}\equiv0 $.
\section{Wave-Particle Interaction}
\label{sec:kinetic:waveparticle}
The kinetic equation of SEP propagation and acceleration includes pitch-angle scattering, which plays a crucial role.
Specifically, according to DSA, 
during gradual SEP events
particle acceleration occurs near the Sun  at the CME-driven shock waves. 
Fast DSA requires that particles experience frequent scattering
back and forth across the shock-wave front.
This scattering may be caused by the turbulence preexisting in the solar wind or, unless the shock wave is
entirely perpendicular, it may be enhanced by the \alf waves that are generated by the accelerated particles streaming from the shock \cite[e.g.][]{Bell1978a,Bell1978b,lee83}.
Therefore, a complete SEP model 
needs to be coupled with a realistic model of
\alf turbulence, including the self-excited one, as well as a model
of particle transport in realistic turbulent 
IMF. 

Within the quasi-linear (QL)
approach, the turbulence is thought of as an ensemble of linear circularly polarized \alf waves with a harmonic electric field:
\be
\label{eq_ecomplex}
\delta\textbf{E}_\perp=(E_x,E_y)=
\delta E\left(
\cos\left(kz-\omega t\right),
\pm\sin\left(kz-\omega t\right)
\right)=
\delta E\left(1, \pm \mathrm{i}\right)e^{-{\rm i}\omega t + {\rm i}kz},
\ee
where $\delta E$ is the field's amplitude.
Hereafter, 
only the real part is implied in complex expressions for real physical quantities.
$z$-axis of the Cartesian coordinate frame, $\left(x,y,z\right)$, is aligned with the magnetic field direction, $\mathbf{b}$.
 Herewith,
 while the frequency, $\omega$, is always positive,
 the wave number, $k$, 
 is positive or negative for the wave modes 
 propagating parallel or 
 anti-parallel to the magnetic field respectively.  
 A choice of sign of $\delta E_y$ accounts for the two types of circular polarization. 
 Thus, there are 4 distinct wave modes.
 We denote quantities for measured for each mode with index $\sigma=\overline{1,4}$.
 For example, the phase speed is 
 $V_\sigma=\omega/k$. 
 Note that $V_\sigma$ has the same sign as $k$.
 
 The oscillating magnetic field, $\delta
\textbf{B}_\perp = (\textbf{b} \times \delta \textbf{E}_\perp) /V_\sigma$, may be found from the induction equation, which gives us an expression for the magnetic field amplitude, $\delta B$, in terms of that for the electric field: \be\label{eq_deltabfromdeltae}(\delta B)^2=\frac{(\delta E)^2}{V_\sigma^2},\ee as well as an expression for the Lorentz force:
\be\label{eq_Lorentz}
\mathbf{F}^{(w)}_L=qZ_i\left(\delta\mathbf{E}+\mathbf{v}\times\delta\mathbf{B}\right)=qZ_i\left[\left(1-\frac{v_\|}{V_\sigma}\right)\delta\mathbf{E}+\frac{\left(\mathbf{v}_\perp\cdot\delta \mathbf{E}\right)}{V_\sigma}\mathbf{b}\right],\ee
$qZ_i$ being the ion charge and $q=|q|$ being the elementary charge. 
The effect of this force on a distribution function $F(\mathbf{R},\mathbf{p}_\perp,p_\|,t)$ is described by the term $\mathbf{F}_L\cdot\frac{\partial F}{\partial \mathbf{p}}$ in the Boltzmann equation,
\be
\label{eq_2_turb}
\mathbf{F}^{(w)}_L\cdot\frac{\partial F}{\partial \mathbf{p}}=qZ_i\delta\mathbf{E}_\perp\cdot\mathbf{G}\{ F \},\qquad\mathbf{G}\{ f \} = \left( 1 - \frac{v_{||}}{V_\sigma} \right) \frac{\partial F}
{\partial \mathbf{p}_\perp} + \frac{ \mathbf{v}_\perp}{V_\sigma} \frac{\partial F}
{\partial p_{||}},
\ee
where the differential operator, $\mathbf{G}\{ F \}$, is only by a numerical factor different from that used in \cite{Ng2003}. The perpendicular components may be expressed terms of the polar angle, $\varphi$: 
\begin{equation}\label{eq_dpperp}
\mathbf{p}_\perp=p_\perp\left( \cos\varphi, \sin\varphi\right),\qquad\frac{\partial f}{\partial \mathbf{p}_\perp}=\frac{\partial f}{\partial p_\perp}\left(\cos\varphi, \sin\varphi\right)+\frac{1}{ p_\perp}\frac{\partial f}{\partial\varphi }\left( -\sin\varphi, \cos\varphi\right).  
\end{equation}

Combining Eqs.~\ref{eq_ecomplex},~\ref{eq_2_turb}~and~\ref{eq_dpperp} we obtain:
\be\label{eq_flcdotdelf}
\mathbf{F}^{(w)}_L\cdot\frac{\partial F}{\partial \mathbf{p}}=qZ_i\delta E\exp(-\mathrm{i}\omega t +\mathrm{i}kz\pm \mathrm{i}\varphi) \left[\left( 1 - \frac{v_{||}}{V_\sigma} \right)\left( \frac{\partial F}
{\partial p_\perp}\pm\frac{\mathrm{i}}{p_\perp}\frac{\partial F}
{\partial \varphi}\right) + \frac{ v_\perp}{V_\sigma} \frac{\partial F}
{\partial p_{||}}\right].\ee
\subsection{Kinetic Response Function}

The perturbed ion distribution function satisfies the Boltzmann equation:
\begin{equation}
\label{eq:boltzmann:turb:full}
\frac{\partial}{\partial t}\left(f+\delta f\right)+
(\mathbf{v}\cdot\nabla)\left(f+\delta f\right)-
\omega_{ci}\frac{\partial}{\partial\varphi}\left(f+\delta f\right) +
\mathbf{F}^{(w)}_L\cdot\frac{\partial}{\partial \mathbf{p}}\left(f+\delta f\right)=0,
\end{equation}
where $\omega_{ci}=\frac{qZ_iB}{m_i}$ is ion-cyclotron frequency,
$\delta f$ is the perturbation of the distribution function 
due to the turbulence.
Naturally, if we keep only the term that are of zeroth order in $\delta E$,
Eq.~\ref{eq:boltzmann:turb:full} yields the equation for the unperturbed
distribution function:
\begin{equation}
\label{eq:boltzmann:turb:0}
\frac{\partial f}{\partial t}+(\mathbf{v}\cdot\nabla)f-\omega_{ci}\frac{\partial f}{\partial\varphi} =0,
\end{equation}
By definition from Section~\ref{sec:basic:distr}, $\frac{\partial f}{\partial\varphi}=0$.
In other words,
the solution of Eq.~\ref{eq:boltzmann:turb:0} is a gyrotropic function, $f(p_\perp,p_\|)$. 
To evaluate both ion scattering by \alf waves and wave excitation one needs to find the perturbation, $\delta f$, of the distribution function.
In the first order approximation, 
$\delta f$ obeys the following equation:
\be
\label{eq:boltzmann:turb:1}
\frac{\partial\delta f}{\partial t}+(\mathbf{v}\cdot\nabla)\delta f-\omega_{ci}\frac{\partial\delta f}{\partial\varphi}=-qZ_i\delta E_\perp\exp(-\mathrm{i}\omega t +\mathrm{i}kz\pm \mathrm{i}\varphi) G\{ f \},
\ee
where: 
\be
\label{eq_2a_turb}
G\{ \psi \} = \left(1 - \frac{ v_{||}}{V_\sigma} \right) \frac{\partial \psi}
{\partial p_\perp} +\frac{ v_\perp}{V_\sigma} \frac{\partial \psi}
{\partial p_{||}}=\sqrt{1-\mu^2}\left[\frac{\partial \psi}{\partial p}+\left(\frac{1}{m_iV_\sigma}-\frac{\mu}p\right)\frac{\partial\psi}{\partial\mu}\right],
\ee 
is a linear differential operator acting on a gyrotropic function $\psi(p_\perp,p_\|)$. 
Eq.~\ref{eq:boltzmann:turb:1} can be solved:
\be
\label{eq_0a_turb}
\delta f =\frac{qZ_i\delta E}{\mathrm{i}\left(\omega-kv_\|\pm\omega_{ci}\right)}\exp(-\mathrm{i}\omega t +\mathrm{i}kz\pm \mathrm{i}\varphi)G\{ f \}.\ee
\subsection{Excitation of Turbulence}
\label{sec:kinetic:wpi:turb}

The self-generated Alfv{\'e}n waves produced in the vicinity of a shock-wave
front have been demonstrated to have important consequences for SEP elemental
abundance variations \cite[]{Ng1999,Tylka1999} and 
the evolution of SEP anisotropies \cite[]{Reames2001}.  

The dispersion relation,
i.e. relation between wave frequency,~$\omega$, and wave number,~$k$, is:
\begin{equation}
\label{eq:dispersion}
\left(\frac{ck}{\omega}\right)^2=\epsilon_{r(l)} (k, \omega),
\end{equation}
where $r$ and $l$ denote right and left polarizations.
Here, $c=\frac1{\sqrt{\mu_0\epsilon_0}}$ is the speed of light, $\epsilon_0,\mu_0$ are the vacuum dielectric and magnetic permeabilities and $\epsilon_{r(l)} (k, \omega)$ 
is a dielectric response function, which is complex in general. Eq.~\ref{eq:dispersion} determines the wave phase speed, $V_\sigma=\omega / k=\pm c / \sqrt{\epsilon_{r(l)}}.$ 
For low frequency \alf waves, $\omega \ll \omega_{ci}$, the dispersion relation gives: 
$\vert V_\sigma\vert = V_A=\sqrt{\frac{B^2}{\mu_0\rho}}$. 
However, for higher frequency $V_\sigma$ depends on both  $\vert k\vert$ and polarization. 
A small wave growth rate due to interaction with energetic ions can be expressed in terms of the small imaginary part of the dielectric response function via Eq.~\ref{eq:dispersion} \citep[see][]{Ichimaru1973}: $-2\Im(\omega)c^2/(\omega V_\sigma^2)=\Im(\epsilon_{r(l)})$. For a harmonic time-dependence for the wave amplitude 
$\propto \exp \left( - \mathrm{i}\omega t\right)=\exp \left[ - i \, \Re(\omega) t
+ \Im (\omega) t \right]$, the wave intensity
($\propto$ square of wave amplitude) grows/decays as $\propto \exp\left[\gamma_{r(l)} (k)t\right],$ where:
$$\gamma_{r(l)} (k) = 2 \Im (\omega)\approx  - \Im
(\epsilon_{r(l)}) \, \omega V_\sigma^2 / c^2.$$
The imaginary part of the dielectric response function can be conveniently expressed in terms of the conductivity, $\epsilon_{r(l)}=1+\frac{\Sigma_{r(l)}}{-\mathrm{i}\omega\epsilon_0},$ so that $\gamma_{r(l)}=-\Re(\Sigma_{r(l)})\mu_0V_\sigma^2.$ The contribution from ions to the conductivity can be found from Eq.~\ref{eq_0a_turb} by calculating the current density, $\mathbf{j}_\perp=qZ_i\int d^3 \mathbf{p}\,\mathbf{v}_\perp\delta f$. Averaging $\delta f$ over $\varphi$ using the easy-to-derive formula, $\frac1{2\pi}\int_0^{2\pi}d\varphi\mathbf{v}_\perp e^{\pm\mathrm{i}\varphi}=\frac{v_\perp}{2}\left(1, \pm\mathrm{i}\right)$, shows that the current is parallel to $\delta \mathbf{E}_\perp$ (see Eq \ref{eq_ecomplex}): $\mathbf{j}_\perp=\Sigma_\perp\delta\mathbf{E}_\perp$, where:
\be
 \Sigma_\perp=\int d^3 \mathbf{p}\frac{q^2Z_i^2}{2\mathrm{i}}\frac{v_\perp G\{f\}}{\omega-kv_\|\pm\omega_{ci}}.
\ee
In agreement with  Eq.~7.59 in \citet{Ichimaru1973}, a formula for the growth rate is as follows:
\bea
\label{eq_1_turb}
\gamma_{r(l)} =  \int
d^3 \mathbf{p}\frac{q^2Z_i^2\mu_0V_\sigma^2}{2} \Im \left( \frac{1}{k v_{||}
- \omega \mp \omega_{ci} - \mathrm{i}{\cdot}0} \right)v_\perp G \{ f \}.
\eea
The pole in Eq.~\ref{eq_1_turb} should be bypassed using the Landau rule
\citep[e.g.][]{Ginzburg1975}, so that
Eq.~\ref{eq_1_turb} can be re-written
in terms of the Dirac $\delta$-function as follows:
\bea
\label{eq_3_turb}
\gamma_\sigma& = &\int d^3 \mathbf{p} \, K_\sigma(\vert k\vert, \mathbf{p})
v_\perp G \{ f \},\\
\label{eq_4_turb}
K_\sigma (\vert k\vert, \textbf{p})& =& \frac{\pi}{2}q^2Z_i^2\mu_0V_\sigma^2 \, \delta
\left(\vert k \vert\left(\mu v-V_\sigma \right)  - g_\sigma\omega_{ci} \right).
\eea
Here, $g_\sigma = \pm 1$, where
index $\sigma=1,2,3,4$ enumerates all combinations of signs of $V_\sigma$ and $g_\sigma$. Note, Eq.~\ref{eq_4_turb} is only valid in the QL approximation, whereas
Eq.~\ref{eq_3_turb} is general and holds true in the non-linear
theory \citep[see][]{Ng2003}. An explicit expression in spherical coordinates reads:
\be\label{eq_5_turb}
\gamma_\sigma=2\pi\int p^2dpd\mu \, K_\sigma(\vert k\vert, \mathbf{p})v\left(1-\mu^2\right)\left[\frac{\partial f}{\partial p}+\left(\frac{1}{m_iV_\sigma}-\frac{\mu}p\right)\frac{\partial f}{\partial\mu}\right]
\ee

\subsection{Particle Scattering}
\label{sec:kinetic:wpi:scat}
The influence of the \alf turbulence on the supra-thermal particles can
be described via  the collision integral, 
in the second order approximation of Eq.~\ref{eq:boltzmann:turb:full} for the gyration-averaged distribution function, $f$:
\be
\label{eq_scat}
\frac{\partial f}{\partial t}+(\mathbf{v}\cdot\nabla)f-\omega_{ci}\frac{\partial f}{\partial\varphi} =\left( \frac{\delta f}{\delta t} \right)_{\rm scat}=-\mathbf{F}^{(w)}_L\cdot\frac{\partial \delta f}{\partial \mathbf{p}},
\ee
where the bi-linear product of rapidly oscillating multipliers should be time-averaged in a usual manner as: $\left( \frac{\delta f}{\delta t} \right)_{\rm scat}=-\frac{qZ_i}{2}\Re\left(\delta \mathbf{E}_\perp^* \cdot \mathbf{G}
\{ \delta f \}\right)$, the superscript asterisk denoting the complex conjugation.  
On substituting $\delta f$ from Eq.~\ref{eq_0a_turb} and using \ref{eq_flcdotdelf} the scattering integral becomes:
\be
\label{eq_0_scat}
\left( \frac{\delta f}{\delta t} \right)_{\rm scat}=G^T\left\{\Re\left(
\frac{q^2Z_i^2(\delta E)^2}{2i\left( kv_\|-\omega\mp\omega_{ci}-i{\cdot}0\right)}
\right)G\{ f \}\right\},
\ee
where 
\bea
G^T\{ \psi \}&=&G\{\psi\}\pm\frac{\left(\mathrm{i}e^{\pm\mathrm{i}\varphi}\right)^*}{p_\perp}\left( 1 - \frac{ v_{||}}{V_\sigma} \right)\frac{\partial\left(e^{\pm\mathrm{i}\varphi}\psi\right)}{\partial\varphi}=\left( 1 - \frac{ v_{||}}{V_\sigma} \right)\frac{1}{p_\perp} \frac{\partial (p_\perp\psi)}
{\partial p_\perp}+ \frac{ v_\perp}{V_\sigma} \frac{\partial\psi}
{\partial p_{||}}=\nonumber\\&=&\frac1{p^2}\frac{\partial\left(p^2\sqrt{1-\mu^2} \psi\right)}{\partial p}+\frac{\partial}{\partial\mu}\left[\left(\frac{\sqrt{1-\mu^2}}{m_iV_\sigma}-\frac{\mu\sqrt{1-\mu^2}}p\right)\psi\right]\eea
is another operator acting on a gyrotropic function, $\psi(p_\perp,p_\|)$. This operator differs from $G\{\psi\}$, because $\delta f$ in contrast with $f$ includes $\varphi$-dependent multiplier, $e^{\pm\mathrm{i}\varphi}$. The electric field amplitude may be expressed in terms of that of magnetic field using \ref{eq_deltabfromdeltae}: $\left(\delta E\right)^2=V^2_\sigma(\delta B)^2$.
Since the energy density in the \alf wave is twice that of the magnetic field, $(\delta B)^2 / 2 \mu_0$, the quantity $(\delta B)^2 / \mu_0$ is the energy density in the \alf wave, which for  a turbulent spectrum of harmonics can be represented as the integral over spectrum and sum over four wave branches: $(\delta B)^2/\mu_0=\sum_\sigma \int\limits_0^{\infty}{
d \vert k\vert \, I_\sigma(\vert k\vert)}$. We arrive at the particle scattering operator in the QL approximation: 
\be
\label{eq_1_scat}
\left( \frac{\delta f}{\delta t} \right)_{\rm scat} = \sum_\sigma  G^T 
\left\{\int\limits_0^\infty{
d\vert k\vert \, I_\sigma\left(\vert k\vert\right)
K^\prime_\sigma(\vert k\vert, \mathbf{p})
G \{ f \}}\right\},
\ee
where the contribution from the pole can be again expressed in terms of the Dirac function:
\be
\label{eq_1a_scat}
K^\prime_\sigma (\vert k\vert, \textbf{p}) = \frac{\pi}{2}q^2Z_i^2\mu_0V_\sigma^2 \, \delta
\left(\vert k \vert\left(\mu v-V_\sigma \right)  - g_\sigma\omega_{ci} \right).
\ee
The kernel $K^\prime_{\sigma}$ in Eq.~\ref{eq_1_scat} appears to be identical to $K_\sigma$ 
in Eq.~\ref{eq_4_turb}.
Such relation is to be expected since the kernels
must be coupled due to
energy conservation in the system comprising 
all ions and all waves 
\citep[see][]{Ng2003}. The latter can be verified using a remarkable conjugation property:
\be\label{eq_conjugation}
\int d^3\mathbf{p}\psi_1G\{\psi_2\}=-\int d^3\mathbf{p}\psi_2G^T\{\psi_1\},\ee
which is true for any pair of gyrotropic functions $\psi_1$ and $\psi_2$. Using \ref{eq_1_scat},\ref{eq_conjugation} and an easy-to-check identity, $G \{ {\cal E}(p) \} = 
v_\perp$,  one can express the total ion energy growth rate 
due to scattering: 
$$\int{d^3 \textbf{p} \, {\cal E}(p) \left( \frac{\delta f}{\delta t} \right)_{\rm scat}}=-\sum_\sigma  \int{d^3\mathbf{p}v_\perp 
\int\limits_0^\infty{
d \vert k\vert \, I_\sigma\left(\vert k \vert\right)
K^\prime_\sigma(\vert k\vert, \mathbf{p})
G \{ f \}}},$$
${\cal E}(p)$ being the ion energy, relativistic in general case. In turn, the total wave energy growth rate may be formulated in terms of $\gamma_\sigma$ from
Eq.~\ref{eq_3_turb}:
$$
\sum_\sigma \int{d\vert k\vert \, I_\sigma\left(\vert k\vert\right) \gamma_\sigma\left(\vert k\vert\right)}=\sum_\sigma \int\limits_0^\infty{
d \vert k\vert \, I_\sigma\left(\vert k \vert\right)\int{d^3\mathbf{p}v_\perp 
K_\sigma(\vert k\vert, \mathbf{p})
G \{ f \}}}
$$
The total energy conservation is controled by the following equation:
\bea
\label{eq_2_scat}
\sum_\sigma \int{d \vert k\vert \, I_\sigma\left(\vert k\vert\right) \gamma_\sigma\left(\vert k\vert\right)}
 + \int{d^3 \textbf{p} \, {\cal E}(p) \left( \frac{\delta f}{\delta t} \right)_{\rm scat}} = 0,
\eea
which holds as long as the
following relation between the kernels of the integrals is fulfilled:
\be
\label{eq_3_scat}
K_\sigma (\vert k\vert, \textbf{p})= K^\prime_\sigma(\vert k\vert, \textbf{p}).
\ee
Note, Eq.~\ref{eq_3_scat} is much more general than any existing
model for wave generation and/or particle scattering.  For comparison, a
similar formula was obtained by \cite{Ng2003} using the non-linear growth rate for
\alf turbulence. As long as Eq.~\ref{eq_3_scat} is true, the momentum conservation can be also proven based upon another easy-to-check identity, $G\left\{p_\|\right\}=\frac{v_\perp}{V_\sigma}$:
\bea
\label{eq_8_scat}
\sum_\sigma \int{d \vert k\vert \, \frac{I_\sigma\left(\vert k\vert\right)}{V_\sigma} \gamma_\sigma\left(\vert k\vert\right)}
 + \int{d^3 \textbf{p} \, p_\| \left( \frac{\delta f}{\delta t} \right)_{\rm scat}} = 0,
\eea
Using the conjugation property Eq.~\ref{eq_conjugation} one can also prove that the isotropic part of the distribution function, $f_0({\cal E})$ always contributes to  dissipation rather than excitation for all \alf wave branches as long as $df_0/d{\cal E}<0$.

The expression for the collision
integral for waves of a given branch becomes very simple and easy to compute in a special frame of reference moving with the phase speed of the wave. A transformed particle velocity in this frame of reference is $\mathbf{v}_\sigma = \mathbf{v} - V_\sigma \mathbf{b}$, so that the  distribution function, $f(\mathbf{R},p_\perp,p_\|, t)$,
transforms as follows: $f(\mathbf{R},p_\perp,p_{\|\sigma}+m_iV_\sigma, t)$.
Under this transformation,
the differential operator, $G\{f\}$, 
involves only the derivative with respect to $\mu_\sigma$:
\begin{equation}
\label{eq_8_turb}
G\left\{f\right\} = 
\frac{\sqrt{1-\mu^2_\sigma}}{m_iV_\sigma}
\frac{\partial f}{\partial\mu_\sigma},
\end{equation}
being taken at constant $p_\sigma$. Eq.~\ref{eq_3_turb} in this frame of reference reads:
\bea
\label{eq_7_turb}
\gamma_\sigma = 2 \pi \int p_\sigma^2 d p_\sigma d \mu_\sigma \frac{\pi\omega_{ci}^2}{2\left(B^2/\mu_0\right)}\delta
\left(\vert k \vert\mu_\sigma v_\sigma  - g_\sigma\omega_{ci} \right) \left[\left( 1 - \mu_\sigma^2 \right) m_iV_\sigma
\frac{\partial f}{\partial \mu_\sigma}
\right].
\eea

If each term in the collision integral is calculated in frame of reference moving with the \alf corresponding \alf wave, the use of Eq.~\ref{eq_8_turb}
gives:
\be
\label{eq_4_scat}
\left( \frac{\delta f}{\delta t} \right)_{scat} = \sum_\sigma
\frac{\partial}{\partial \mu_\sigma} \left( D^\sigma_{\mu \mu}
\frac{\partial f}{\partial \mu_\sigma} \right),
\ee
where
\bea
\label{eq_5_scat}
D^\sigma_{\mu \mu} = \frac{\pi\omega_{ci}^2}{2(B^2/\mu_0)}(1 - \mu_\sigma^2) 
\int\limits_0^\infty d k \, I_\sigma(k)
\delta
\left(\vert k \vert\mu_\sigma v_\sigma - g_\sigma\omega_{ci} \right).
\eea
In the QL approximation, the integral by $k$ can be taken using the
presence of $\delta$-function in $K_\sigma$ (see Eq.~\ref{eq_4_turb}).
Thus, for given $\sigma$ and $\mu_\sigma$, Eq.~\ref{eq_5_scat} becomes \citep[][]{jokipii66,lee82,lee83}:
\be
\label{eq_6_scat}
D^\sigma_{\mu \mu} =\frac{\pi\omega_{ci}}{2 (B^2/\mu_0)}\left(1-\mu_\sigma^2\right) \frac{\omega_{ci} }{v_\sigma \vert \mu_\sigma \vert}I_\sigma \left( \frac{\omega_{ci}}{v_\sigma \vert \mu_\sigma \vert}
\right),
\ee
for two wave branches with  $g_\sigma = \textrm{sign}(\mu_\sigma)$, while for the two other branches $D^\sigma_{\mu \mu}$ vanishes. 
\section{Application to the diffusive limit}
. 
Since both types of diffusion, spatial and pitch-angle, 
are different representations of the same physical process,
scattering on the magnetic field irregularities, the spatial diffusion coefficient along
the magnetic field, $D_{xx}$, is expressed in terms of
 $D_{\mu\mu}$ 
\citep{jokipii66,earl74b}:
\begin{equation}
D_{xx}=
\frac{v^2}{8}\int\limits_{-1}^1\frac{\left(1-\mu^2\right)^2}{D_{\mu\mu}}d\mu
\end{equation}
In turn, the pitch-angle diffusion coefficient, $D_{\mu\mu}$, 
may be expressed in terms of the \alf wave turbulence spectrum, 
as discussed in Section~\ref{sec:kinetic:wpi:scat}.
In the QL approximation, using Eqs.~\ref{eq_4_scat} and~\ref{eq_6_scat}
one can obtain a closed form of Eq.~\ref{eq_6_DSA} for the spatial diffusion coefficient in the diffusive approximation:
\be
\label{eq_7a_DSA}
D_{xx} = \frac{v^3B^2}{2 \pi\mu_0 \omega_{ci}^2}
\int_{-1}^{1} \frac{\left( 1 - \mu^2 \right) \vert \mu \vert}
{I_-\left( \frac{\omega_{ci}}{v\vert \mu \vert}\right)+I_+\left( \frac{\omega_{ci}}{v\vert \mu \vert}\right)} d \mu.
\ee

For the two wave branches contributing to 
Eq.~\ref{eq_6_scat}, 
the propagation directions are opposite for each $\mu$. 
We can assume that in the \alf wave turbulence the left and right polarized waves are balanced and that their total wave energy for a given $k$ equals $I_+(k)$ for waves propagating along the field direction ( $V_\sigma=+V_A$) and $I_-(\vert k\vert)$ for the oppositely propagating waves ($V_\sigma=-V_A$). One can notice that under this assumption for any positive $\mu$, hence, for a given $g_\sigma=1$, the contribution to  Eq.~\ref{eq_6_scat} is proportional $\frac12\left[I_-\left( \frac{\omega_{ci}}{v \vert\mu\vert}
\right)+I_+\left( \frac{\omega_{ci}}{v \vert\mu\vert}
\right)\right]$, i.e. the half of the total wave spectral energy, while the other half would contribute to scattering the particles with the negative $\mu$. Herewith, we consider only the high-energy particles with $v_\sigma\gg V_A$ and thus neglect the difference between $v_\sigma$ and $v$, which allows us to write the total scattering rate as follows:
\be
\label{eq_7_scat}
D_{\mu \mu} =\frac{\pi\omega_{ci}}{4 (B^2/\mu_0)}\left(1-\mu^2\right) \frac{\omega_{ci} }{v \vert \mu \vert}\left[I_-\left( \frac{\omega_{ci}}{v\vert \mu \vert}\right)+I_+\left( \frac{\omega_{ci}}{v\vert \mu \vert}\right)\right],
\ee 
In terms of an integral over the turbulence spectrum, the spatial diffusion
coefficient can be written as:
\be
\label{eq_7b_DSA}
D_{xx} = \frac{v B^2}{ \pi\mu_0}\int_{k_r}^\infty
\frac{d k \left( k^2 - k_r^2 \right)}{k^5\left[I_-(k)+I_+(k)\right]},
\ee
where the resonant wave number, $k_r$, is the inverse of the Larmor
radius, i.e., $k_r = {e Z_i B / p}$, and $k=\frac{\omega_{ci}}{v\vert \mu \vert}=\frac{k_r}\mu$.

One can also use Eq.~\ref{eq_3_DSA} to evaluate the pitch-angle dependence
of the distribution function in the expression for the wave growth rate
(see Eq.~\ref{eq_7_turb}).
  Again, by neglecting the difference between
$\mu$ and $\mu_\sigma$, one obtains:
\bea
\label{eq_8_DSA}
\gamma_\sigma = 2 \pi \int d p d \mu p^2 (1 - \mu^2)
\left( - k v^2 \frac{1 - \mu^2}{2 D_{\mu
\mu}} \frac{\partial f_0} {\partial s} \right) K_\sigma(k, \omega,
\textbf{p}).
\eea
In the QL limit, using Eq.~\ref{eq_6_scat} this becomes:
\bea
\label{eq_9_DSA}
\gamma_\sigma = -\frac{\pi V_\sigma}{\vert k \vert \left(I_+(k)+I_-(k)\right)} \int_{p_{res}(k)}^\infty d p p^3  
\frac{p_{res} (k)}{m_i} \left( 1 - \frac{p^2_{res}(k)}
{p^2} \right) \frac{\partial f_0}{\partial s},
\eea
where the resonant value of momentum, $p_{res}$, for a given $k$, is
defined as $p_{res}(k) = m_i \omega_{ci} / k$.
\section{Kolmogorov's Spectrum of Turbulence}
Some further evaluations can be performed, 
if one assumes the Kolmogorov's spectrum for turbulence:
$I_-(k)\propto k^{-5/3}$, $I_+(k)\propto k^{-5/3}$, at $k>k_0$. We take the total spectrum to be
\begin{equation}\label{eq_Kolm_1}I_-(k)+I_+(k)=\frac{I_C}{k^{5/3}},
\end{equation}
the parameter $I_C$ characterizes the turbulence level 
and is specified below. 
In this section, we calculate both the scattering rate,
$D_{\mu\mu}$, and the spatial diffusion coefficient, $D_{xx}$,
for this kind of turbulence spectrum.  
Eq. \ref{eq_7_scat} yields the following scattering rate:
\be
\label{eq_Kolm_2_a}
D_{\mu \mu} =\frac{v}{\lambda_{\mu\mu}}\left(1-\mu^2\right) \vert \mu \vert^{2/3},\qquad\lambda_{\mu\mu}=\frac{4}\pi\frac{  B^2/\mu_0}{ I_C}r_L^{1/3} ,
\ee
$r_L=v/\omega_{ci}$ being the Larmor radius and $\lambda_{\mu\mu}$ being the characteristic value of the mean free path with respect to pitch-angle scattering. 

An alternative and more consistent way to parameterize the turbulence level is to take into account an energy integral. 
By assuming, as stated above, 
a negligible level of turbulence below some minimum wave number, 
i.e. at $k\le k_0$, which correspond to large spatial scales, 
we floor an integration span by condition, $k\ge k_0$: 
\begin{equation}
    w_-+w_+=\frac{\left(\delta B\right)^2}{\mu_0}  
    =\int_{k_0}^\infty
d k\left[I_-(k)+I_+(k)\right]=\frac32I_Ck_0^{-2/3}.
\end{equation}
In this way, the mean free path can be expressed in terms 
of the turbulent energy density and $k_0$:
\begin{equation}
\lambda_{\mu\mu}=\frac6\pi\frac{B^2}{\left(\delta B\right)^2}
\frac{r_L^{1/3}}{k_0^{2/3}},\qquad \left(\delta B\right)^2=\mu_0\left(w_-+w_+\right).
\end{equation}
In agreement with the Bohm-like estimate, $\lambda\sim\frac{B^2}{(\delta B)^2}r_L$, 
i.e. mean free path is proportional to a (large) factor, $\frac{B^2}{(\delta B)^2}$. 
However, the distinction is in a different dependence on the particle momentum (via the Larmor radius): $\propto p^{1/3}$ with the derived formula versus $\propto p$ in the Bohm-like estimate.

The spatial diffusion coefficient, $D_{xx}$, in terms of the energy spectrum of turbulence is given by Eq.~\ref{eq_7b_DSA}. It may be also expressed in terms of the mean free path, $\lambda_{xx}$: 
\begin{equation}
\label{eq_Kolm_2a}
D_{xx}=\frac13\lambda_{xx} v,\qquad
\lambda_{xx} = \frac{3 B^2}{ \pi\mu_0}\int_{k_r}^\infty
\frac{d k \left( k^2 - k_r^2 \right)}{k^5\left[I_-(k)+I_+(k)\right]},
\end{equation}
where $k_r(p)=\frac{eZ_iB}p$ is the inverse of the Larmor radius. With ansatz (\ref{eq_Kolm_1}) this mean free path is only by a numerical factor different from
above introduced $\lambda_{\mu\mu}$ and equals:
\begin{equation}
\label{eq_Kolm_3}
\lambda_{xx} 
=\frac{54}{7\pi}\frac{B^2/\mu_0}{I_C} r_L^{1/3}.
\end{equation}

Particularly, one can choose $I_C$ in such way, that the mean free path estimate Eq.~\ref{eq_Kolm_3} would agree with that provided by \cite{li03}, which had been also used by \cite{Sokolov2004}:
\begin{equation}\label{Gang_Li}
    \lambda_{xx}=\lambda_0\frac{R}{1AU}\left(\frac{pc}{1GeV}\right)^{1/3},
\end{equation}
where $\lambda_0\sim0.1\div0.4$~AU is a free parameter. 
The same dependence on the particle momentum, 
but a different dependence, $\lambda_{xx}\propto (R/1AU)^{2/3}$,
on the heliocentric distance was assumed by \cite{Zank2007}. 
The mean free path in Eq.~\ref{Gang_Li} corresponds to the choice of $I_C$ as follows:
\begin{equation}\label{Gang_Li_a}
I_C=\frac{54B^2}{7\pi\mu_0\lambda_0\frac{R}{1AU}} r_{L0}^{1/3},
\end{equation}
$r_{L0}=\frac{1GeV}{ceB}$ being the Larmor radius for the particle momentum $1GeV/c$. 
\cite{Sokolov2009} employed the Kolmogorov spectrum with 
$I_C$ from Eq.~\ref{Gang_Li_a} to provide a seeding level of 
the \alf wave turbulence upstream the shock wave, 
which is strongly enhanced by the SEPs accelerated by the DSA mechanism and up-streaming the shock. 
Far upstream, the turbulence is not affected by the SEP of low intensity, so that the mean free path as in Eq.~\ref{eq_Kolm_2a} with $I_C$ from Eq.~\ref{Gang_Li_a} correspond to the estimate in Eq.~\ref{Gang_Li}. 
We see that with the use of the Kolmogorov's spectrum of turbulence, 
the dependence of mean free path on the particle momentum $\lambda\propto p^{1/3}$ is achieved which can be found in literature and the spatial modulation of the turbulence spectrum may be applied to achieve a desired spatial modulation of the mean free path. 

On the other hand, by expressing the mean free path in terms of  $k_0$ and $(\delta B)^2$
\begin{equation}
\lambda_{xx}=  \frac{81}{7\pi}\frac{B^2}{\left(\delta B\right)^2}\frac{r_L^{1/3}}{k_0^{2/3}}\equiv \frac{81}{7\pi}\frac{B^2}{\left(\delta B\right)^2}\frac{r_{L0}^{1/3}}{k_0^{2/3}}\left(\frac{pc}{1GeV}\right)^{1/3},
\end{equation}
in the last identity we separated the momentum-dependent factor, same as in \citep[]{li03,Sokolov2004,Zank2007,Sokolov2009}. At the same time we keep the dependence on large Bohm-like factor $\frac{B^2}{(\delta B)^2}=\frac{B^2}{\mu_0\left(w_-+w_+\right)}$, which can be consistently obtained from the turbulence-driven model for IH and SC.

To close the model, we need the estimate for $k_0$. 
As the first trial of the model in 
\citet{boro18}, we performed simulations with $$k_0={\rm const}\sim0.1/R_S.$$ 
However, more realistic seems to be an observation-based constraint for the maximum spatial scale in the turbulence, $L_{\rm max}$, which relates to the minimum wave vector and scales about linearly with the heliocentric distance: 
$$k_0^{-1}=\frac{L_{\rm max}(R)}{2\pi},\qquad L_{\rm max}(R)\sim 0.03R,$$
so that, on evaluating the numerical factor $\frac{81}{7\pi(2\pi)^{2/3}}\approx0.92$ we arrive at the following formulae :
\begin{equation}
\lambda_{xx}   \approx0.9\frac{B^2/\mu_0}{w_-+w_+}\left(L_{\rm max}^2r_{L0}\right)^{1/3}\left(\frac{pc}{1GeV}\right)^{1/3},\qquad D_{xx}=\frac13\lambda_{xx}v
\end{equation}
and, with another factor of $14/27$,
\begin{equation}
\lambda_{\mu\mu}   \approx0.5\frac{B^2/\mu_0}{w_-+w_+}\left(L_{\rm max}^2r_{L0}\right)^{1/3}\left(\frac{pc}{1GeV}\right)^{1/3}, \qquad D_{\mu\mu}=\frac{v}{\lambda_{\mu\mu}}\left(1-\mu^2\right)|\mu|^{2/3}
\end{equation}

\section{Numerical implementation}
\label{sec:numerics}
\subsection{M-FLAMPA}
\label{sec:mflampa}

To solve the Parker Eq.~\ref{eq:parker}, \cite{boro18} developed the Multiple Field Line Advection Model for Particle Acceleration (M-FLAMPA). 
M-FLAMPA is based on the method first proposed by \cite{Sokolov2004} and 
reduces a 3-D problem of particle propagation in the IMF
to a multitude 
of much simpler 1-D problems of the particle transport along a single line of the Interplanetary Magnetic Field (IMF).

We choose the Lagrangian coordinates for a given fluid element equal to the Eulerian coordinates of this element at the initial time instant, ${\bf R}_L={\bf R}|_{t=0}$. For numerical simulations, the initial grid is chosen as follows. Let the points, $\left({\bf R}_{0}^{l\lambda}\right)|_{t=0}$, form a grid on a segment of a spherical heliocentric surface of the radius of $R=2.5\,R_\odot$. 
The indices $l,\lambda$ enumerate both this spherical grid's points and the magnetic field lines passing through these points.
For each $l,\lambda$ one can solve numerically for the field line passing through the point $\left({\bf R}_{0}^{l\lambda}\right)|_{t=0}$ by solving the following ordinary differential equation with the boundary condition:
\begin{equation}
\label{eq:dxds}
\frac{d\mathbf{R}^{l\lambda}(s)}{ds}=\left(\mathbf{b}\left(\mathbf{R}^{l\lambda}(s),t\right)\right)_{t=0}, 
\qquad
\mathbf{R}^{l\lambda}(0)=\left({\bf R}^{l\lambda}_{0}\right)_{t=0},\end{equation}
where $s$ is the curve length  along the magnetic field line.
When all the lines are constructed, one can introduce a grid $s_i$ along each line. 
Now, the choice of the grid in Lagrangian coordinates, $\left({\bf R}^{l\lambda}_{i}\right)_L= {\bf R}^{l\lambda}(s_i)$, ensures that for fixed $l,\lambda$ all points with different $i$ initially belong to the magnetic field line. 
Then, one can numerically solve the multitude of ordinary differential equations, Eq.~\ref{eq:DxDt}, 
to trace the spatial location for all Lagrangian grid points 
in the evolving fluid velocity field, ${\bf u}({\bf R},t)$, 
as long as the latter is known. Since the magnetic field lines are frozen into a moving plasma, still all the grid points with fixed $l,\lambda$ belong to the same magnetic field line and the kinetic equation for these points is independent and effectively one-dimensional in space.
In this way, the three-dimensional kinetic equation for waves reduces to a two-dimensional multitude of one-dimensional equations.

\appendix

\section{Focused Transport Equation  and Single Particle Dynamics}

The kinetic treatment of SEP
was used throughout the present paper.
In other words, the particle population 
and its properties were encapsulated
in the distribution function.
However, there exists another approach, 
which suggests
solving equations of single particle dynamics directly for a relatively small yet representative set of particles.
In this appendix we want to emphasize a deep
connection between these very different approaches by revealing how
equations of single particle dynamics are 
actually a part of the kinetic equation
governing the particle population.

We reproduce the focused transport equation below (see Section~\ref{sec:kinetic:FTE}):
\bea
\label{app:eq:FTE}
\frac{Df}{Dt} &+& 
v\mu\frac{\partial f}{\partial s} +\left[
\frac13\frac{D\ln\rho}{Dt}+
\frac{1-3\mu^2}{6}\frac{D\ln \left(B^3/\rho^2\right)}{Dt}
-\frac{\mu}{v}\mathbf{b}\cdot\frac{D\mathbf{u}}{Dt}
\right]p\frac{\partial f}{\partial p}+\nonumber\\
&+& \frac{1-\mu^2}{2}\left[
-v\frac{\partial\ln B}{\partial s} +
\mu\frac{D\ln\left(\rho^2/B^3\right)}{Dt}- 
\frac{2}{v}\mathbf{b}\cdot\frac{D\mathbf{u}}{Dt}
\right]\frac{\partial f}{\partial\mu}=
\left(\frac{\delta f}{\delta t}\right)_{\rm scat}
\eea

We introduce the parallel, $p_{||} = \mu \, p$, and perpendicular, $p_\perp
= (1 - \mu^2)^{1/2} \, p$, components of the momentum instead of pitch-angle, $\mu$. 
On substituting 
$\frac{\partial f}{\partial \ln p}=p_\perp\frac{\partial f}{\partial p_\perp}+p_\|\frac{\partial f}{\partial p_\|}$
and 
$\frac{\partial f}{\partial \mu}=-\frac{p\mu}{\left(1-\mu^2\right)^{1/2}}\frac{\partial f}{\partial p_\perp}+p\frac{\partial f}{\partial p_\|}$, one can rewrite Eq.~\ref{app:eq:FTE} in the form:
\bea
\label{app:eq:FTE2}
\frac{D f}{D t} + \left(\frac{d s}{d t}\right)_p \frac{\partial f} {\partial s} + \left(\frac{dp_\perp}{d t}\right)_p \frac{\partial
f}{\partial p_\perp}  + \left(\frac{d p_{||}}
{d t} \right)_p \frac{\partial f} {\partial p_{||}} = \left(
\frac{\delta f}{\delta t} \right)_{\rm scat}.
\eea
Here, the coefficients $(\frac{d\,}{dt})_p$ in the kinetic equation are
time-derivatives of canonical variables of a particle along its trajectory and 
are given by the following equations \citep[cf.][]{northrop1963}:
\bea
\label{app:eq:spd1}
\left(\frac{d s}{d t} \right)_p\:&=&\: v_{||}\\
\label{app:eq:spd2}
\left(\frac{d p_\perp}{d t} \right)_p&=&\:\frac12 \left(\frac{D
\ln{B}}{D t} + \frac{\partial \ln{B}}{ \partial s} v_{||} \right)
p_\perp=\frac12\left(\frac{d \ln B}{d t} \right)_pp_\perp\\
\label{app:eq:spd3}
\left(\frac{d p_{||}}{ d t} \right)_p&=&\,- \frac{p_\perp^2}{2m_i} \frac{\partial \ln{B}}{
\partial s} + \frac{D \ln(\rho/B)}{ D t} p_{||} -  m_i \textbf{b}
\cdot \frac{D \textbf{u}}{ D t}
\eea

Here, $m_i=\frac{p}{v}$ is the ion mass, which in application to relativistic particles should be substituted with the relativistic mass $\sqrt{m^2_i+p^2/c^2}=\frac{m_i}{\sqrt{1-v^2/c^2}}$ . 
The terms in Eqs.~\ref{app:eq:spd1}-\ref{app:eq:spd3} have simple and 
straightforward physical meaning.
Thus, we see that evolution of the distribution
function in Eq.~\ref{app:eq:FTE2} is
governed by: (1) particle's guiding center displacement along the field line, see Eq.~\ref{app:eq:spd1}; 
(2) conservation of the magnetic moment, $p_\perp^2 / (2 m_i B)$, see Eq.~\ref{app:eq:spd2}; 
(3) magnetic mirror force, see first term on
RHS of Eq.~\ref{app:eq:spd3}; 
(4) first-order Fermi acceleration, with the
conservation of another adiabatic invariant, $p_{||} \delta s$ (clear if rewritten as $-\frac{D \ln{\delta s}}{ D t} p_{||}$), see the second term on RHS of Eq.~\ref{app:eq:spd3}); 
(5) action of a non-inertial
force $\propto - D \textbf{u} / D t$, see Eq.~\ref{app:eq:spd3}; 
and (6) particle scattering and sources
(RHS of Eq.~\ref{app:eq:FTE2}). 
Regarding process (3), the term in Eq.~\ref{app:eq:spd3} is the force repelling the particle from a magnetic mirror.
 For a time-independent magnetic field (i.e., $D B / D t = 0$),
the action of this force balances the energy change due to the perpendicular momentum increase (adiabatic focusing), thus ensuring the energy conservation.

The above Eqs.~\ref{app:eq:spd1}-\ref{app:eq:spd3} are convenient for computations
using particle methods, especially within the Monte-Carlo approach. 
Similarly, the $\mu$-dependent form of Eq.~\ref{app:eq:FTE} can be also be solved in this way by integrating the equations for $\left(\frac{d\ln p}{ d t} \right)_p$ and $\left(\frac{d \mu}{d t} \right)_p$ with the RHSs being the factors by $\frac{\partial f}{\partial\ln p}$ and $\frac{\partial f}{ \partial \mu}$ terms in Eq.~\ref{app:eq:FTE}.

\section{Quasi-linear perturbation of distribution function}
\label{app:ql}

Electro-magnetic fields of \alf waves exert Lorentz force on ions
in solar wind.
The effect on the distribution function is described by the term 
in the Boltzmann equation:
\begin{equation}
    \mathbf{F}_L^{(w)}\cdot\frac{\partial F}{\partial\mathbf{p}}=
    qZ\delta\mathbf{E}_\perp\cdot\mathbf{G}\{F\}
\end{equation}
with the differential operator
\begin{equation}
    \mathbf{G}\{F\}=
    \left(1-\frac{v_\|}{V_\sigma}\right)\frac{\partial F}{\partial\mathbf{p}_\perp}+
    \frac{\mathbf{v}_\perp}{V_\sigma}\frac{\partial F}{\partial p_\|}
\end{equation}

A number of relations involving $\mathbf{G}$ are used
in derivations in Section~\ref{sec:kinetic:waveparticle}.

Operator $\mathbf{G}$ can be expressed in several ways.
The perpendicular components may be expressed terms of the polar angle, $\varphi$: 
\begin{equation}\label{eq_dpperp}
\mathbf{p}_\perp=p_\perp\left( \cos\varphi, \sin\varphi\right),\qquad\frac{\partial f}{\partial \mathbf{p}_\perp}=\frac{\partial f}{\partial p_\perp}\left(\cos\varphi, \sin\varphi\right)+\frac{1}{ p_\perp}\frac{\partial f}{\partial\varphi }\left( -\sin\varphi, \cos\varphi\right).  
\end{equation}

Then
\begin{align}
    \mathbf{G}\{F\}&=
    \left(1-\frac{v_\|}{V_\sigma}\right)
    \left(
    \frac{\partial F}{\partial p_\perp} \mathbf{e}_\perp+
    \frac{1}{p_\perp}\frac{\partial F}{\partial\varphi}\mathbf{e}_\varphi
    \right)+
    \frac{v_\perp}{V_\sigma}\frac{\partial F}{\partial p_\|}\mathbf{e}_\perp\nonumber\\
   &= G\{F\}\,\mathbf{e}_\perp+
    \left(1-\frac{v_\|}{V_\sigma}\right)
    \frac{1}{p_\perp}\frac{\partial F}{\partial\varphi}\mathbf{e}_\varphi
\end{align}
Here we introduced a new scalar differential operator
\begin{equation}
   G\{ \psi \} = \left(1 - \frac{ v_{||}}{V_\sigma} \right) \frac{\partial \psi}
{\partial p_\perp} +\frac{ v_\perp}{V_\sigma} \frac{\partial \psi}
{\partial p_{||}}
\end{equation}

If we express $p_\perp$ and $p_\|$ in terms of
magnitude of momentum, $p$, and 
cosine of pitch-angle, $\mu$, 
we obtain the following expression:
\begin{equation}
    G\{ \psi \} = \sqrt{1-\mu^2}\left[\frac{\partial \psi}{\partial p}+\left(\frac{1}{m_iV_\sigma}-\frac{\mu}p\right)\frac{\partial\psi}{\partial\mu}\right],
\end{equation}

In the course of determining the effect of waves
on particle scattering, the following
bilinear form appears:
\begin{align}
    \mathbf{F}_L^{(w)}\cdot\frac{\partial\delta f}{\partial\mathbf{p}}&=
    qZ_i\Re\left[\delta\mathbf{E}_\perp\right]\cdot\mathbf{G}\{\delta f\}\nonumber\\
    &=
    \left(qZ_i\right)^2\Re\left[\delta\mathbf{E}_\perp\right]\cdot
    \mathbf{G}\left\{
    \frac{\mathbf{G}\{f\}}{i\left(\omega-kv_\|\pm\omega_{ci}\right)}\cdot\Re\left[\delta\mathbf{E}_\perp\right]
    \right\}\nonumber\\
    &=
    \Re\left[\delta\mathbf{E}_\perp\right]\cdot\mathfrak{A}\cdot\Re\left[\delta\mathbf{E}_\perp\right]
\end{align}
Note that here we explicitly stated that only the real part
of the electric field is used in the expression.
The equation that contains this bilinear form
is averaged over period of rapid oscillations of the vector
$\delta\mathbf{E}_\perp$.
The following holds:
\begin{align}
    \langle\delta\mathbf{E}^*_\perp\cdot\mathfrak{A}\cdot\delta\mathbf{E}_\perp\rangle = 
    &\langle\Re\left[\textbf{}\mathbf{E}_\perp\right]\cdot\mathfrak{A}\cdot\Re\left[\delta\mathbf{E}_\perp\right]\rangle+
     \langle\Im\left[\textbf{}\mathbf{E}_\perp\right]\cdot\mathfrak{A}\cdot\Im\left[\delta\mathbf{E}_\perp\right]\rangle+\nonumber\\
     &i\left(
      \langle\Re\left[\textbf{}\mathbf{E}_\perp\right]\cdot\mathfrak{A}\cdot\Im\left[\delta\mathbf{E}_\perp\right]\rangle-
       \langle\Im\left[\textbf{}\mathbf{E}_\perp\right]\cdot\mathfrak{A}\cdot\Re\left[\delta\mathbf{E}_\perp\right]\rangle\right)\nonumber\\
       =&2 \langle\Re\left[\textbf{}\mathbf{E}_\perp\right]\cdot\mathfrak{A}\cdot\Re\left[\delta\mathbf{E}_\perp\right]\rangle,
\end{align}
where the last transition is possible due
to $\delta\mathbf{E}_\perp$ being a simple harmonic field.
Thus, we have:
\begin{align}
    \mathbf{F}_L^{(w)}\cdot\frac{\partial\delta f}{\partial\mathbf{p}}&= 
    \frac{qZ_i}2\big\langle \delta\mathbf{E}^*_\perp\cdot\mathbf{G}\{\delta f\}\big\rangle\nonumber\\
    &=\frac{qZ_i}2\big\langle\delta\mathbf{E}^*_\perp\cdot\left(
    G\{\delta f\}\mathbf{e}_\perp+\left(1-\frac{v_\|}{V_\sigma}\right)\frac{1}{p_\perp}\frac{\partial\delta f}{\partial\varphi}\mathbf{e}_\varphi
    \right)\big\rangle\nonumber\\
    &=\frac{\left(qZ_i\delta E\right)^2}2...
\end{align}

A remarkable conjugation property:
\be\label{eq_conjugation}
\int d^3\mathbf{p}\psi_1G\{\psi_2\}=-\int d^3\mathbf{p}\psi_2G^T\{\psi_1\},\ee
which is true for any pair of gyrotropic functions $\psi_1$ and $\psi_2$.

\bibliographystyle{authyear}
\bibliography{wholeilr_June8}
\clearpage
\end{document}